\documentclass[aps,12pt,showpacs,showkeys,nofootinbib]{revtex4-1}

\usepackage[english]{babel}
\usepackage{amsmath}
\usepackage{amssymb}
\usepackage{amsbsy}
\usepackage{amstext}
\usepackage{graphicx}
\usepackage{pst-node}
\usepackage{verbatim}
\usepackage{braket}
\usepackage{tikz}
\usepackage{ulem}
\usetikzlibrary{decorations.pathmorphing}
\newcommand{\be}{\begin{eqnarray}}
\newcommand{\ee}{\end{eqnarray}}
\newcommand{\bdm}{\begin{displaymath}}
\newcommand{\edm}{\end{displaymath}}
\newcommand{\ds}{\displaystyle}
\newcommand{\ba}{\begin{array}}
\newcommand{\ea}{\end{array}}
\newcommand{\pa}[1]{\left(#1\right)}
\newcommand{\paq}[1]{\left[#1\right]}

\newcommand{\K}{{\mathbf k}}
\newcommand{\Q}{{\mathbf q}}

\newcommand{\intkq}{\int_{\K\,,\Q}}
\newcommand{\eps}{\epsilon}

\begin{document}
\normalem %This is necessary to recover the standard behaviour of \emph{} under ulem package.
\title{Classical Gravitational Self-Energy from Double Copy}

\author{Gabriel Luz Almeida$^{\rm 1}$, Stefano Foffa$^{\rm 2}$ and Riccardo Sturani$^{\rm 3}$}

\affiliation{$(1)$ Departamento de F\'\i sica Te\'orica e Experimental, Universidade Federal do Rio Grande do Norte, Natal-RN 59072-970, Brazil\\
  $(2)$ D\'epartement de Physique Th\'eorique and Center for Astroparticle
  Physics, Universit\'e de Gen\`eve, CH-1211 Geneva, Switzerland\\
  $(3)$ International Institute of Physics, Universidade Federal do Rio Grande
  do Norte, Campus Universit\'ario, Lagoa Nova, Natal-RN 59078-970, Brazil}

\email{gabriel.luz@fisica.ufrn.br, stefano.foffa@unige.ch, riccardo@iip.ufrn.br}

\begin{abstract}
  We apply the classical double copy to the calculation of self-energy of
  composite systems with multipolar coupling to gravitational field, obtaining
  next-to-leading order results in the gravitational coupling
  $G_N$ by generalizing color to kinematics replacement rules known in
  literature.
  When applied to the multipolar description of the two-body system, the
  self-energy diagrams studied in this work correspond
  to tail processes, whose physical interpretation is of 
  radiation being emitted by the non-relativistic source, scattered by the
  curvature generated by the binary system and
  then re-absorbed by the same source.
  These processes contribute to the conservative two-body dynamics
  and the present work represents a decisive step towards the systematic
  use of double copy within the multipolar post-Minkowskian expansion.
\end{abstract}



\maketitle

\section{Introduction}
Links between the gauge and gravity theories first appeared in scattering
amplitude computations, as first shown within a string theory context
by the Kawai-Lewellen-Tye identities \cite{Kawai:1985xq} relating tree
level closed and open string amplitudes, later extended to a correspondence
between S matrix elements in gauge theory and gravity \cite{BjerrumBohr:2003vy}.
More recently the Bern-Carrasco-Johansson (BCJ) formalism \cite{Bern:2008qj}
has provided a general mechanism for viewing gravitons as double copies of
gluons at perturbative level.

The BCJ relations state that squaring non-Abelian Yang-Mills amplitudes in generic dimension $d$, and applying a set of rules to map color into
kinematics degrees of freedom, gravitational amplitudes are recovered, which
however do not coincide with General Relativity but include a scalar,
\emph{dilaton} field and a 2-form gauge field $B_{\mu\nu}$. The BCJ double copy
has been verified in a variety of supersymmetric field theories,
see \cite{Carrasco:2015iwa,Bern:2019prr} for reviews. 

A remarkable application of double copy to non-perturbative classical solutions
in Yang-Mills theory on one side, and Kerr-Schild black holes in General Relativity on the other, was shown in \cite{Monteiro:2014cda,Kim:2019jwm}, and
further development on
the classical side were made in \cite{Goldberger:2016iau}, where the \emph{long-distance} radiation gluon field emitted by a set of gauge charges has been
computed and mapped into asymptotic radiation field in a theory of gravity
plus a dilaton.
This latter result has been extended in \cite{Goldberger:2017ogt,Li:2018qap} to the case of
spinning particles, in \cite{Shen:2018ebu} to next-to-leading
order in coupling (in the post-Minkowskian regime of gravity) and in
\cite{Goldberger:2019xef} to next-to-leading order with finite-size sources of non-zero spin, see also
\cite{Bautista:2019tdr} for double copy application to gravitational radiation and spin effects.

For other relevant work on the classical double copy: see \cite{Plefka:2018dpa} for application to the two-body effective
gravitational potential in the post-Newtonian approximation, with possible
problems arising at $O(G^2_N)$ with respect to leading order \cite{Plefka:2019hmz},
and the seminal work
\cite{Bern:2019nnu,Bern:2019crd} for the determination of the two-body potential at third post-Minkowskian order.

In the present work we show the computation of self-energy diagrams representing \emph{forward scattering}
of non-relativistic
sources described by their multipolar coupling to gauge and gravity fields to next-to-leading order
in the gauge/gravity coupling, by extending previously derived rules for gauge
charge/kinematic variable duality. According to standard post-Newtonian (PN) approximation
to General Relativity \cite{Blanchet:2013haa}, this processes contribute to the conservative two-body
dynamics starting at 4PN order.

In the post-Newtonian approach to the two-body dynamics it is customary to separate the \emph{near} from the \emph{far} zone.
In the former the interactions
between the binary constituents are mediated by the constrained, non-radiative
longitudinal modes of gravity, in the latter
gravitational radiative degrees of freedom are also relevant and the source
is modeled as a single object with multipoles.
The real part of self-energy diagram amplitudes in the far zone complements
the near zone derivation of the effective two-body dynamics \cite{Foffa:2019rdf,Foffa:2019yfl}, while the
imaginary part relates via the optical theorem to the radiated energy.

The paper is structured as follows: in sec.~\ref{sec:method} we introduce the double copy method
applied to source coupled to gauge fields and gravity in the multipole expansion.
In sec.~\ref{sec:self_energy_lo} we give the details of the correspondence, verifying the matching
of the ``square'' of the gauge self-energy amplitude with the General Relativity plus dilatonic and axionic amplitude, checking the correspondence at next-to-leading order in gauge/gravitational coupling in sec.~\ref{sec:self_energy_nlo}.
We finally conclude in sec.~\ref{sec:discussion}.

\section{Method}
\label{sec:method}
We show how the mapping between the square of gauge amplitudes and gravity ones work
in the case of multipole-expanded sources.
On the gauge side we consider the bulk action\footnote{We adopt the mostly plus signature for the metric,
  i.e. Minkowski metric $\eta_{\mu\nu}=\rm{diag}(-1,1,\ldots,1)$.}
\be
{\cal S}^{(gauge)}_{bulk}=-\int {\rm d}^{d+1}x\paq{\frac 14F^a_{\mu\nu}F^{a\mu\nu}+\frac 12\pa{\partial_\mu A^{a\mu}}^2}
\ee
in terms of the field strength $F^a_{\mu\nu}$ with structure constant $f^{abc}$, where we have displayed explicitly the Feynman gauge fixing term in terms of the gauge
field $A_\mu^a$ (resulting in the
standard propagator $P[A^a_\mu,A^b_\nu]=-i\delta^{ab}\eta_{\mu\nu}/\pa{\K^2-k_0^2}$, boldface character denoting 3-vectors),
and a system of
classical, spinning Yang-Mills color charges coupled to gluons, described
by a trajectory $x^\mu$, a color charge $c_a$ and a spin $S^{\mu\nu}$ (all three
depending on the world-line parameter $\tau$), whose dynamics is described by the
world-line action  summed over particles
\be
\label{eq:wl_gauge}
\ba{rcl}
   \ds{\cal S}^{(gauge)}_{wl}&=&\ds \sum_{p\in parts} g\int {\rm d}x^\mu c_{ap} A^a_\mu
     -\frac\kappa 2 \int {\rm d}\tau c_{ap}S^{\mu\nu}_pF^a_{\mu\nu}\\
   &\simeq&\ds g\int {\rm d}t\pa{q_a A^a_0+d_a^i F^a_{i0}+
     \frac 12Q_a^{ij}F^a_{i0,j}-\frac 12\pa{\mu_{ak}+\kappa c_aS_k}\epsilon^{kij} F^a_{ij}
     +\ldots}\,,
\ea
\ee
where in the second line we moved from the description in terms of fundamental
constituents to the one in terms of an extended object with multipoles.\footnote{Greek indices run over $d+1$ space-time dimensions, Latin indices
$i,j\ldots$ over space coordinates only, Latin indices $a,b,\ldots,h$ run gauge
color indices, $q^a$ is the gauge charge $q_a\equiv \sum_p c_a(\tau(t))$, $d^i_a$ the electric dipole $d^i_a\equiv\sum_p c_a x_p^i$,
$Q^{ij}_a\equiv \sum_p c_ax^i_px^j_p$ is the electric quadrupole,
$\mu_{ak}$ the magnetic dipole
$\mu^a_k\equiv \frac 12\epsilon_{kij}\sum_p c_a\frac{dx^i_p}{dt}x^j_p$ and
$\kappa$ is a numerical coefficient determining the strength of the
chromomagnetic interaction.}
The spin anti-symmetric tensor $S^{\mu\nu}$ has 6
components, we then adopt a \emph{spin supplementary condition}
\cite{HANSON1974498} to reduce them to the three physical
ones, implying that $S^{i0}\sim S^{ij}v^j$ (and $S^k\equiv \frac 12\epsilon^{kij}S_{ij}$
in eq.~(\ref{eq:wl_gauge})).

On the gravity side we have that the degrees of freedom are represented by the
metric $g_{\mu\nu}$, the dilaton $\psi$ and the axion $B_{\mu\nu}$ with field
strength $H_{\mu\nu\rho}$ defined by
\be
H_{\mu\nu\rho}\equiv
\partial_{\mu}B_{\nu\rho}+\partial_{\rho}B_{\mu\nu}+\partial_{\nu}B_{\rho\mu}\,.
\ee
The gauge-fixed bulk action is\footnote{Note that the dilaton is not canonical normalized
  here, it (and $B_{\mu\nu}$) has canonical dimensions, as the metric
  fields $\phi,A_i,\sigma_{ij}$ that will be introduced in eq.~(\ref{eq:met_KK}).}
\be
\label{eq:bulk}
   {\cal S}_{bulk}^{(gda)}=\int {\rm d}^{d+1}x\sqrt{-g}\paq{2\Lambda^2\pa{R-\frac 12\Gamma^\mu\Gamma_\mu}
     -2(d-1)(\partial\psi)^2-\frac 16 e^{-\frac{4\psi}\Lambda}H_{\mu\nu\rho}H^{\mu\nu\rho}-\pa{\partial_\mu B^{\mu\nu}}^2}\,,
\ee
where $\Lambda\equiv (32\pi G_N)^{-1/2}$ (it has dimension of
$\sqrt{{\rm mass}/{\rm length}}$ in $d=3$)
and $\Gamma^\mu\equiv\Gamma^{\mu}_{\nu\rho}g^{\nu\rho}$.
The world-line action is
\be
\label{eq:mult}
\ba{rcl}
\ds{\cal S}_{wl}^{gda}&=&\ds\sum_{p\in parts} -m_p\int_p {\rm d}\tau\,e^{\frac\psi\Lambda}-\frac12 \int_p {\rm d}\tau S^{\mu\nu}\Omega_{\mu\nu}
+\frac 1{4\Lambda}\int_p {\rm d}x^\rho H_{\rho\mu\nu}S_p^{\mu\nu}\\
   &\simeq&\ds\int {\rm d}t \left\{\frac 12 Eh_{00}+\frac 12\epsilon^{ijk}L_ih_{0j,k}
     +\frac 12I^{ij}{\cal E}_{ij}+\frac 16 I^{ijk}{\cal E}_{ij,k}+
     \frac 23 J^{ij}{\cal B}_{ij}+\ldots\right.\\
     &&\ds\left. \quad+\frac 1{\Lambda}\paq{-E\psi-\frac 12\pa{I^{ij}\psi_{,ij}-I\ddot\psi}+\ldots
     +\frac 14S^{ij}H_{0ij}-\frac 13J^{ij}\epsilon_{ikl}H_{0jk,l}+\ldots}\right\}\,,
\ea
\ee
   where the world-line coupling (inclusive of the angular velocity tensor $\Omega_{\mu\nu}$ \cite{HANSON1974498}) have been expanded in multipoles for an ensemble of particles or equivalently for a finite
   size source with small internal velocities: $h_{\mu\nu}\equiv g_{\mu\nu}-\eta_{\mu\nu}$ is the gravitational perturbation around the Minkowski metric $\eta_{\mu\nu}$,
   $E$ is the total energy of the
   source, $L^i\equiv\eps^{ijk}S_{jk}$ its total angular momentum (dual to the spin anti-symmetric tensor, whose mixed time-space polarization $S^{0i}$ vanish in the
   source center of mass),
   $I^{ij}$ and $I^{ijk}$ are respectively the traceless mass quadrupole and octupole,
   $J^{ij}$ the magnetic quadrupole
   $J^{ij}\equiv\frac 12\int {\rm d}^dx\, x^kT^{0l}\pa{\epsilon^{ikl}x^j+\epsilon^{jkl}x^i}$,
   and the eletric ${\cal E}_{ij}$ and magnetic ${\cal B}_{ij}$ parts of the
   Riemann tensor $R^\mu_{\ \nu\rho\sigma}$ in the source rest frame take, respectively, the form
   ${\cal E}_{ij}=R_{0i0j}$, ${\cal B}_{ij}=\frac 12\epsilon_{ikl}R_{0jkl}$ (see app.~\ref{app:details} for details).

 By  generalizing the double copy rules to the coupling of gluons to source
   multipole moments we will express the ``square'' of classical self-energy
   diagrams on the gauge side as classical self-energy diagrams in the
   gravitational theory, to $O(G_N)$ interactions beyond the leading order
   diagrams.

\section{Self-energy diagrams at leading order}
\label{sec:self_energy_lo}
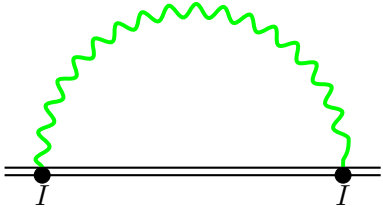
\begin{figure}
  \begin{center}
    \begin{tikzpicture}
      \draw [black, thick] (0,0) -- (5,0);
      \draw [black, thick] (0,-0.1) -- (5,-0.1);
      \draw[decorate, decoration=snake, line width=1.5pt, green] (0.5,0)  arc (180:0:2.);
      \filldraw[black] (0.5,-0.1) circle (3pt) node[anchor=north] {$I$};
      \filldraw[black] (4.5,-0.1) circle (3pt) node[anchor=north] {$I$};
    \end{tikzpicture}
  \end{center}
  \caption{Self-energy diagram for a generic radiative multipole source $I$. The green
    wavy field represents the gauge/gravity interaction, while the black double line
    stands for the composite source.}
  \label{fig:self_energy_lo}
\end{figure}

We compute in this section self-energy diagrams like the one in fig.~\ref{fig:self_energy_lo},
with generic multipole $I$ insertion at the extended object world-line.

We will compute quantities in space-dimension $d=3$  but it will be helpful to
keep $d$ generic in the computations to check that explicit dependence on
  $d$ cancels in the sum of gravitational, dilatonic and
axionic effective actions, as it happens in scattering amplitudes \cite{Scherk:1974mc,Bern:1999ji}.

In our non-relativistic setup we find convenient to use the Kaluza-Klein
parametrization of the metric \cite{Kol:2007bc}\footnote{We adopt the same
  symbol $A^a_i$ for the gauge field and $A_i$ for the mixed time-space
  component of gravity, the former being accompanied by the gauge index should
  avoid confusion between the two.}
\be
\label{eq:met_KK}
g_{\mu\nu}=e^{2\phi/\Lambda}\pa{
  \ba{cc}
  -1 & \dfrac{A_i}\Lambda\\
  \dfrac{A_j}\Lambda & \quad e^{-c_d\phi/\Lambda}\pa{\delta_{ij}+\dfrac{\sigma_{ij}}\Lambda}-
  \dfrac{A_iA_j}{\Lambda^2}
  \ea}\,,
\ee
where $c_d\equiv 2(d-1)/(d-2)$ and $d$ is the number of purely space dimensions.
Such decomposition has the virtue to provide diagonal propagators for the gravity
fields, which we list below together with the dilaton and axion propagators
\be
\label{eq:propsKK}
\left.
\ba{rcl}
\ds P[\phi,\phi]&=&\ds-\frac 1{2c_d}\\
\ds P[A_i,A_j]&=&\ds \frac{\delta_{ij}}2\\
\ds P[\sigma_{ij},\sigma_{kl}]&=&\ds -\frac 12\pa{\delta_{ik}\delta_{jl}+\delta_{il}\delta_{jk}-\frac 2{d-2}\delta_{ij}\delta_{kl}}\\
\ds P[\psi,\psi]&=&\ds-\frac 1{4(d-1)}\\
\ds P[B_{\mu\nu},B_{\rho\sigma}]&=&\ds-\frac 12\pa{
  \eta_{\mu\rho}\eta_{\nu\sigma}-\eta_{\mu\sigma}\eta_{\nu\rho}}
\ea
\right\}\times \frac i{\K^2-k_0^2-i\epsilon}\,.
\ee
The Lagrangian for the bulk fields up to cubic interactions is reported in
eq.~(\ref{eq:sbulk_cubic}).

Below, we compute self-energy diagrams which, at leading order in $G_N$ ($g$),
are due to processes represented by the diagram in fig.~\ref{fig:self_energy_lo}.
As it will be clear in the following, all such diagrams involve time
derivatives
of the source, hence the lowest non-vanishing on the gravity (gauge) side
involves electric quadrupole (dipole) moments.
The Green's function involved in the process is the Feynman one, which is the
correct prescription for self-energy diagrams \cite{Foffa:2019eeb}.

\subsection{Electric moments}
\label{ssec:QQ}
The lowest order non-vanishing self-energy diagram involve two quadrupole
sources and in the gravity side receives contributions from exchange of
gravitational and dilatonic modes\footnote{We adopt the notation
  $\int_\K\equiv\int \frac{{\rm d}^dk}{(2\pi)^d}$}:
\be
\label{eq:QQ_GR}
\ba{rcl}
\ds {\cal S}^{(I^2)}_{GR}&=&\ds -\frac 1{8\Lambda^2}\int\frac{{\rm d}k_0}{2\pi} I^{ij}(k_0)I^{kl}(-k_0)
\int_\K\frac 1{\K^2-k_0^2}\left\{
-\frac 18k_0^4\pa{\delta_{ik}\delta_{jl}+\delta_{il}\delta_{jk}-\frac 2{d-2}\delta_{ij}\delta_{kl}}\right.\\
&&\left.\ds\qquad+\frac{k_0^2}4\pa{k_ik_k\delta_{jl}+k_jk_l\delta_{ik}}-\frac 1{2c_d}\pa{k_ik_j+\frac{k_0^2\delta_{ij}}{d-2}}
\pa{k_kk_l+\frac{k_0^2\delta_{kl}}{d-2}}\right\}\,,
\ea
\ee
\be
\label{eq:QQ_psi}
{\cal S}_{\psi}^{(I^2)}=\frac 1{32(d-1)\Lambda^2}\int \frac{{\rm d}k_0}{2\pi}I^{ij}(k_0)I^{kl}(-k_0)
\int_\K\frac 1{\K^2-k_0^2}\pa{k_ik_j-k_0^2\delta_{ij}}\pa{k_kk_l-k_0^2\delta_{kl}}\,,
\ee
and no contribution from the axion, as it does not couple to electric moments.
Notably the sum of gravitational (\ref{eq:QQ_GR}) and dilatonic (\ref{eq:QQ_psi}) amplitudes display a factorizable structure, where terms explicitly dependent on the number
of space dimensions $d$ cancel
\be
\label{eq:QQ_GRpsi}
   {\cal S}_{GR+\psi}^{(I^2)}=\frac 1{32\Lambda^2}\int \frac{{\rm d}k_0}{2\pi}
   I^{ij}(k_0)I^{kl}(-k_0)\int_\K\frac 1{\K^2-k_0^2}
   \pa{k_ik_k-k_0^2\delta_{ik}}\pa{k_jk_l-k_0^2\delta_{jl}}\,.
\ee
On the gauge side the electric dipole self-energy process gives
\be
\label{eq:dd_A}
{\cal S}_A^{(d^2)}&=&\frac{g^2}2\int\frac{{\rm d}k_0}{2\pi}d^{a,i}(k_0)d^{b,k}(-k_0)
\int_\K\frac 1{\K^2-k_0^2} \langle F_{\K 0i}^aF_{-\K 0k}^b\rangle'\,,
\ee
where primed brackets $\langle\cdots\rangle'$ stands for field Green's functions
stripped of factors $-i/(\K^2-k_0^2)$ and delta functions for each propagator,
e.g. $\int_\Q\langle A^a_{\K\mu}A^b_{\Q\nu}\rangle=-i/(\K^2-k_0^2)\langle A^a_{\K\mu}A^b_{-\K\nu}\rangle'$.

Following standard procedure, we apply the substitutions\footnote{
  The $g\to 1/(2\Lambda)$ agrees with eq.~(58) of \cite{Goldberger:2016iau},
  where $d$ denotes the number of space-time dimensions.} $g\to 1/(2\Lambda)$
and promote the gauge color indices to \emph{space} index to ``square''
the integrand of eq.~(\ref{eq:dd_A}) according to the rule
\be
\label{eq:rule_d}
\ba{l}
\ds d^{a,i}\to \frac{I^{ij}}2\,,\\
\ds \langle F^a_{\K 0i}F^b_{-\K 0k}\rangle'=
\pa{k_0^2\delta_{ik}-k_ik_k}\delta^{ab}\to (k_0^2\delta_{ik}-k_ik_k)(k_0^2\delta_{jl}-k_jk_l)\,.
\ea
\ee
One then obtains
\be
\label{eq:QQ_DC}
 {\cal S}_A^{(d^2)}\to {\cal S}_{DC}^{(I^2)}=\frac 1{32\Lambda^2}\int
 \frac{{\rm d}k_0}{2\pi}I^{ij}(k_0)I^{kl}(-k_0)
 \int_\K\frac 1{\K^2-k_0^2}\pa{k_0^2\delta_{ik}-k_ik_k}\pa{k_0^2\delta_{jl}-k_jk_l}\,,
\ee
which equals the sum of eqs.~(\ref{eq:QQ_GR}) and (\ref{eq:QQ_psi}) given in eq.~(\ref{eq:QQ_GRpsi}).

The above results can be straightforwardly generalized to higher order $2^{r+2}$-th electric moments $I^{iji_1\ldots i_r}$ for gravity
\be
\label{eq:OO_GR}
\ba{rcl}
\ds {\cal S}^{(I^2_{r+2})}_{GR}&=&\ds -\frac 1{2\paq{\pa{r+2}!}^2\Lambda^2}
\int\frac{{\rm d}k_0}{2\pi} I^{iji_1\cdots i_r}(k_0)I^{klk_1\cdots k_r}(-k_0)
\int_\K\frac{k_{i_1}\cdots k_{i_r}k_{k_1}\cdots k_{k_r}}{\K^2-k_0^2}\\
&&\ds\times\left\{
-\frac 18k_0^4\pa{\delta_{ik}\delta_{jl}+\delta_{il}\delta_{jk}-\frac 2{d-2}\delta_{ij}\delta_{kl}}\right.\\
&&\left.\ds\qquad+\frac{k_0^2}4\pa{k_ik_k\delta_{jl}+k_jk_l\delta_{ik}}-\frac 1{2c_d}\pa{k_ik_j+\frac{k_0^2\delta_{ij}}{d-2}}
\pa{k_kk_l+\frac{k_0^2\delta_{kl}}{d-2}}\right\}\,,
\ea
\ee
for the dilaton
\be
\label{eq:OO_psi}
\ba{rcl}
   \ds{\cal S}_{\psi}^{(I^2_{r+2})}&=&\ds\frac 1{8(d-1)\paq{\pa{r+2}!}^2\Lambda^2}
   \int \frac{{\rm d}k_0}{2\pi}I^{iji_1\cdots i_r}(k_0)I^{klk_1\cdots k_r}(-k_0)\\
   &&\ds\times  \int_\K\frac{k_{i_1}\cdots k_{i_r}k_{k_1}\cdots k_{k_r}}{\K^2-k_0^2}
   \pa{k_ik_j-k_0^2\delta_{ij}}\pa{k_kk_l-k_0^2\delta_{kl}}\,,
\ea
\ee
and for the gauge field coupling to the $2^{r+1}$ multipole $d^{a,ii_1\cdots i_r}$
\be
{\cal S}_A^{(d^2_{r+1})}=\frac{g^2}{2\paq{\pa{r+1}!}^2}\int\frac{{\rm d}k_0}{2\pi}
d^{a,ii_1\ldots i_r}(k_0)d^{b,kk_1\ldots k_r}(-k_0)\int_\K
\frac{k_{i_1}\ldots k_{i_r}k_{k_1}\ldots k_{k_r}}{\K^2-k_0^2}
\pa{k_0^2\delta_{ik}-k_ik_k}\,.\nonumber\\
\ee
Applying previous rules (\ref{eq:rule_d}) completed with
\be
d^{a,ii_1\ldots i_r}\to \frac 1{\pa{r+2}}I^{iji_1\ldots i_r}\,,
\ee
the double copy of the gauge electric dipole self-energy can be derived to be
\be
\label{eq:OO_DC}
\ba{rcl}
\ds{\cal S}_{DC}^{(I^2_{r+2})}&=&\ds\frac 1{8\paq{\pa{r+2}!}^2\Lambda^2}\int
\frac{{\rm d}k_0}{2\pi}
I^{iji_1\ldots i_r}(k_0)I^{klk_1\ldots k_r}(-k_0)\int_\K
\frac{k_{i_1}\ldots k_{i_r}k_{k_1}\ldots k_{k_r}}{\K^2-k_0^2}\\
&&\ds\qquad\times\pa{k_ik_k-k_0^2\delta_{ik}}
\pa{k_jk_l-k_0^2\delta_{jl}}\,,
\ea
\ee
which much like in the electric quadrupole case (\ref{eq:QQ_DC}) equates the
sum of (\ref{eq:OO_GR}) and (\ref{eq:OO_psi}).

\subsection{Magnetic moments}
In the magnetic multipole moment case we have from GR
\be
\label{eq:JJ_GR}
\ds{\cal S}_{GR}^{(J^2)}=\ds\frac{\epsilon_{imn}\epsilon_{krs}}{36\Lambda^2}\int\frac {{\rm d}k_0}{2\pi}J^{ij}(k_0)J^{kl}(-k_0)
   \int_\K\frac{k_nk_s}{\K^2-k_0^2}\paq{k_0^2\pa{\delta_{jl}\delta_{mr}+\delta_{jr}\delta_{lm}}-k_jk_l\delta_{mr}}\,,
   \ee
and from the axion
\be
\label{eq:JJ_ax}
{\cal S}_B^{(J^2)}=\ds\frac{\epsilon_{imn}\epsilon_{krs}}{36\Lambda^2}\int\frac {{\rm d}k_0}{2\pi}J^{ij}(k_0)J^{kl}(-k_0)
   \int_\K\frac{k_nk_s}{\K^2-k_0^2}\ds\paq{k_0^2\pa{\delta_{jl}\delta_{mr}-\delta_{jr}\delta_{lm}}-k_jk_l\delta_{mr}}\,,
\ee
with vanishing contribution from the dilaton. Note that in the sum of the
gravitational and axionic contributions the terms where the Levi-Civita tensors
have no indices contracted between themselves cancel, whereas the remaining ones add up.

On the gauge side
\be
\label{eq:mm_A}
      {\cal S}_A^{(\mu^2)}=\frac{g^2}8\epsilon_{imn}\epsilon_{krs}\int
      \frac{{\rm d}k_0}{2\pi}
      \mu^{ai}(k_0)\mu^{bk}(-k_0)\int_\K \frac 1{\K^2-k_0^2} \langle F^a_{\K mn}F^b_{-\K rs}\rangle'\,,
\ee
and making the substitutions
\be
\label{eq:rule_m}
\ba{rl}
&\ds\mu^{ai}\to\frac 23 J^{ij}\,,\\
&\ds\frac 14\epsilon_{imn}\epsilon_{krs}\langle F_{\K mn}^aF^b_{-\K rs}\rangle'=\epsilon_{imn}\epsilon_{krs}k_mk_r\delta_{ns}\delta^{ab}\\
= &\ds (\K^2\delta_{ik}-k_ik_k)\delta^{ab}\to\pa{\K^2\delta_{ik}-k_ik_k}\pa{k_0^2\delta_{jl}-k_jk_l}\,,
\ea
\ee
one has
\be
\label{eq:JJ_DC}
   {\cal S}^{\mu^2}_A\to {\cal S}_{DC}^{J^2}=\frac 1{18\Lambda^2}\int\frac{{\rm d}k_0}{2\pi}
   J^{ij}(k_0)J^{kl}(-k_0)\int_\K\frac 1{\K^2-k_0^2}\pa{\K^2\delta_{ik}-k_ik_k}
   \pa{k_ 0^2\delta_{jl}-k_jk_l}\,,
   \ee
which equates the sum of eqs.~(\ref{eq:JJ_GR}) and (\ref{eq:JJ_ax}).
Note that to obtain the gravitational magnetic result we did not ``square''
the gauge magnetic dipole result (\ref{eq:mm_A}) but rather combine it
with the electric dipole (\ref{eq:dd_A}), which, beside being justified
\emph{a posteriori} as it gives the expected result, is the correct
prescription for preserving magnetic parity.

Like for the self-energy electric dipole of subsec.~\ref{ssec:QQ}, this result
can be generalized to all magnetic multipole moments $J^{iji_1\ldots i_r}$, for
standard gravity
\be
\label{eq:MM_GR}
\ba{rcl}
\ds{\cal S}_{GR}^{(J^2_{r+2})}&=&\ds
\frac {1}{9\paq{\pa{r+2}!}^2\Lambda^2}\int \frac{{\rm d}k_0}{2\pi}
J^{iji_1\dots i_r}(k_0)\,J^{klk_1\cdots k_r}(-k_0)\int_{\K}\frac {k_{i_1}\cdots k_{i_r}k_{k_1}\dots k_{k_r}}{\K^2-k_0^2}\\
&&\ds\paq{(\delta_{ik}\K^2-k_ik_k)(\delta_{jl}k_0^2-k_j k_l)-k_0^2 \epsilon_{ikn}\epsilon_{jls}k_n k_s}\,,
\ea
\ee
for the axion
\be
\label{eq:MM_ax}
\ba{rcl}
\ds{\cal S}_B^{(J^2_{r+2})}&=&\ds\frac 1{9\paq{\pa{r+2}!}^2\Lambda^2}
\int\frac{{\rm d}k_0}{2\pi}J^{iji_1\cdots i_r}(k_0)J^{klk_1\cdots k_r}(-k_0)
\frac{k_{i_1}\cdots k_{i_r}k_{k_1}\cdots k_{k_r}}{\K^2-k_0^2}\\
&&\ds\times\paq{(\delta_{ik}\K^2-k_ik_k)(\delta_{jl}k_0^2-k_j k_l)+
  k_0^2 \epsilon_{ikn}\epsilon_{jls}k_n k_s}\,,
\ea
\ee
and vanishing contribution from the dilaton.

On the gauge side one has
\be
   {\cal S}_A^{(\mu^2_{r+1})}=\frac{g^2\epsilon_{imn}\epsilon_{krs}}{8\paq{\pa{r+1}!}^2}\int\frac{{\rm d}k_0}{2\pi}
   \mu^{a,ii_1\cdots i_r} \mu^{b,kk_1\cdots k_r}
   \int_\K\frac{k_{i_1}\cdots k_{i_r}k_{k_1}\cdots k_{k_r}}{\K^2-k_0^2}
   \pa{\K^2\delta_{ik}-k_ik_k}\delta^{ab},
\ee
which using the double copy rules (\ref{eq:rule_d}) and (\ref{eq:rule_m}), complemented with
\be
\mu^{a,ii_1\ldots i_r}\to\frac 4{3\pa{r+2}}J^{iji_1\cdots i_r}\,,
\ee
one obtains
\be
\ba{rcl}
   {\cal S}_{DC}^{(J^2_{r+2})}&=&\ds\frac 2{9\paq{\pa{r+2}!}^2\Lambda^2}\int\frac{{\rm d}k_0}{2\pi}J^{iji_1\ldots i_r}(k_0)J^{klj_1\ldots j_r}(-k_0)\\
   &&\ds\times\int_\K\frac{k_{i_1}\ldots k_{i_r}k_{k_1}\ldots k_{k_r}}{\K^2-k_0^2}
   \pa{\K^2\delta_{ik}-k_ik_k}\pa{k_0^2\delta_{jl}-k_jk_l}\,,
\ea  
\ee
equalling the sum of (\ref{eq:MM_GR}) and (\ref{eq:MM_ax}) much like in the
magnetic quadrupole case (\ref{eq:JJ_DC}).

\section{Self-energy diagrams at next-to-leading order}
\label{sec:self_energy_nlo}

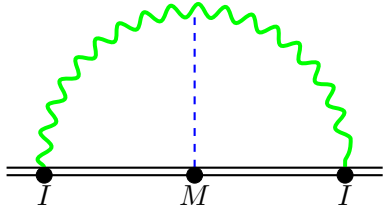
\begin{figure}
  \begin{center}
    \begin{tikzpicture}
      \draw [black, thick] (0,0) -- (5,0);
      \draw [black, thick] (0,-0.1) -- (5,-0.1);
      \draw[decorate, decoration=snake, line width=1.5pt, green] (0.5,0)  arc (180:0:2.);
      \draw [blue, thick, dashed] (2.5,0) -- (2.5,2);
      \filldraw[black] (0.5,-0.1) circle (3pt) node[anchor=north] {$I$};
      \filldraw[black] (4.5,-0.1) circle (3pt) node[anchor=north] {$I$};
      \filldraw[black] (2.5,-0.1) circle (3pt) node[anchor=north] {$M$};
    \end{tikzpicture}
  \end{center}
  \caption{Self-energy diagram of the next-to-leading order process involving the
    scattering of modes emitted by a radiative multipole source $I$ onto
    a mode generated by a generic multipole $M$.}
  \label{fig:self_energy_nlo}
\end{figure}

Having succeeded in the warm-up exercise of double-copying the self-energy
processes without bulk interaction, we now move to the less trivial self-energy
at next order in $G_N$ order, i.e. $O(G_N^2)$ which involves one cubic
interaction in the bulk, see fig.\ref{fig:self_energy_nlo}.
One can distinguish two cases
according to which type of source the third gravitational mode is attached to:
a conserved multipole (like energy and angular momentum) or a radiative multipole.
We will treat here the former case in which the additional world-line insertion
has a total energy $E$ vertex, also known as \emph{tail} process \cite{BD88}, as the
back-scattering induced by the gravitational longitudinal mode sourced by the
total energy induces radiation ``tails'' propagating inside the light cone.

\subsection{Tail diagrams with electric moments}
The pure gravitational process involving electric quadrupoles
gives an effective action \cite{Foffa:2011np}
\be
\label{eq:EQQ_GR}
\ba{rcl}
\ds {\cal S}_{GR}^{(EI^2)}&=&\ds
\frac{E}{16\Lambda^4}\int\frac{{\rm d}k_0}{2\pi}I^{ij}(k_0)I^{kl}(-k_0)
\int_{\K,\Q}\frac 1{\K^2-k_0^2}\frac 1{(\K+\Q)^2-k_0^2}\frac 1{\Q^2}\\
&&\ds\times\left\{\frac{k_0^6}8\pa{\delta_{ik}\delta_{jl}+\delta_{il}\delta_{jk}-\frac {2\delta_{ij}\delta_{kl}}{d-1}}\right.\\
&&\ds \quad+\frac{k_0^4}2\left(2\delta_{jl}k_i q_k+\frac{1}{2(d-1)}\left\{\paq{(k+q)_k (k+q)_l+2q_kq_l}\delta_{ij}+\delta_{kl}k_i k_j\right\}\right)\\
  &&\ds \quad +
\frac{k_0^2}2\bigg(\frac{(d-2)}{2(d-1)}k_i k_j\paq{(k+q)_k (k+q)_l+2q_kq_l}\\
&&\ds\qquad\qquad +k_j\pa{k+q}_l\paq{k_k q_i-k_iq_k-\K\cdot(\K+\Q)\delta_{ik}}
\Big)\Big\}\,,
\ea
\ee
and the the dilaton contribution
\be
\label{eq:EQQ_psi}
\ba{rcl}
\ds {\cal S}^{(EI^2)}_{\psi}&=&\ds
\frac{E}{64(d-1)\Lambda^4}\int\frac{{\rm d}k_0}{2\pi}I^{ij}(k_0)I^{kl}(-k_0)
\int_{\K,\Q}\frac 1{\K^2-k_0^2}\frac 1{(\K+\Q)^2-k_0^2}\frac 1{\Q^2}\\
&&\ds\times\left\{k_0^6\delta_{ij}\delta_{kl}
-k_0^4\left\{\paq{(k+q)_k (k+q)_l+2q_kq_l}\delta_{ij}+\delta_{kl}k_i k_j\right\}\right.\\
&&\ds\qquad\left.+k_0^2k_i k_j\pa{\pa{k+q}_k\pa{k+q}_l+2q_kq_l}\right\}\,,
\ea
\ee
with no contribution from the axion.
The sum of the integrands of the gravitational and dilatonic contributions,
eqs.(\ref{eq:EQQ_GR}) and (\ref{eq:EQQ_psi}), turns out to be independent of $d$
and can be recast in a ``perfect square'' form
with the addition of (non-trivially) vanishing terms
\be
\label{eq:MEE_DC}
\ba{rcl}
\ds{\cal S}_{GR+\psi}^{(EI^2)}&=&\ds\frac E{64\Lambda^4}\int\frac{{\rm d}k_0}{2\pi}
I^{ij}(k_0)I^{kl}(-k_0)\int_{\K,\Q}
\frac 1{\K^2-k_0^2}\frac 1{(\K+\Q)^2-k_0^2}\frac 1{\Q^2}\\
&&\ds \times k_0^2\Big\{\pa{k_0^2\delta_{ik}-\pa{k+q}_ik_k-q_iq_k}\pa{k_0^2\delta_{jl}-\pa{k+q}_jk_l-q_jq_l}\\
&&\ds \quad +\delta_{jl}\paq{2k_0^2\pa{q_ik_k+q_k \pa{k+q}_i}-
k_i\pa{k+q}_k\pa{\pa{\K^2-k_0^2}+\pa{\pa{\K+\Q}^2-k_0^2}-\Q^2}}\\
&&\quad+\paq{k_ik_jq_kq_l-q_iq_j\pa{k+q}_k\pa{k+q}_l}\Big\}\,,
\ea
\ee
which is suggestive of the double-copy structure, as the last two lines of
eq.~(\ref{eq:MEE_DC}) can be shown to identically vanish, see
app.~\ref{sec:dc_show} for details.

The analog process on the gauge side, i.e. electric dipole self-energy at $O(g^2)$ with respect to leading order with conserved charge insertion, contributes
to the effective action according to

\be
\label{eq:qdd_A}
\ba{rcl}
\ds {\cal S}_{A}^{(qd^2)}&=&\ds
i\frac{g^4}2q^a\int\frac{{\rm d}k_0}{2\pi}d^b_i(k_0)d^c_{k}(-k_0) \pa{i f^{def}}
\int_{\K,\Q}\frac 1{\K^2-k_0^2}\frac 1{(\K+\Q)^2-k_0^2}\frac 1{\Q^2}\\
&&\ds \langle\pa{A_0^aF^b_{\K 0i}F_{-\K-\Q 0k}^c}\pa{\partial_\mu A_\nu^dA^e_\rho A^f_\sigma}\rangle'
\eta^{\mu\rho}\eta^{\mu\sigma}\\
&=&\ds-g^4q^a\int\frac{{\rm d}k_0}{2\pi}d^b_i(k_0)d^c_{k}(-k_0) \pa{i f^{abc}}
\int_{\K,\Q}\frac 1{\K^2-k_0^2}\frac 1{(\K+\Q)^2-k_0^2}\frac 1{\Q^2}\\
&&\ds\qquad \times k_0\paq{k_0^2\delta_{ik}-\pa{k+q}_ik_k-q_iq_k}\,,
\ea
\ee
where in the first passage Green's functions are not to be taken between
  fields within the same parenthesis and we adopted a mixed direct-Fourier space
  notation.

Using the gauge-gravity mapping rules derived in the previous section,
completed with \cite{Bern:2008qj}
\be
\label{eq:rule_fabs}
\ds if^{abc}\to\Gamma^{\mu\nu\rho}(k_1,k_2,k_3)\equiv\frac 12\paq{\eta^{\mu\nu}\pa{k_1-k_2}^\rho+\eta^{\rho\mu}\pa{k_3-k_1}^\nu+\eta^{\nu\rho}\pa{k_2-k_3}^\mu}\,,
\ee
and
\be
\label{eq:rule_qdd}
\ba{l}
\ds\langle F_{\K 0i}^aA_\mu^b\rangle'=i\pa{k_0\eta_{i\mu}-k_i\eta_{0\mu}}\delta^{ab}\to
-\pa{k_0\eta_{i\mu}-k_i\eta_{0\mu}}\pa{k_0\eta_{k\mu'}-k_k\eta_{0\mu'}}\,,\\
\ds\langle A_0^aA_\mu^b\rangle'=\eta_{0\mu}\delta^{ab}\to \eta_{0\mu}\eta_{0\rho}\\
J^{a0}\to T^{00}\,,\qquad q^a\to E\,,
\ea
\ee
eq.~(\ref{eq:qdd_A}) can be double-copied into
\be
\ba{rcl}
\ds {\cal S}_{DC}^{(EI^2)}&=&\ds
\frac{E}{64\Lambda^4}\int\frac{{\rm d}k_0}{2\pi}I^{ij}(k_0)I^{kl}(-k_0)
\int_{\K,\Q}\frac 1{\K^2-k_0^2}\frac 1{(\K+\Q)^2-k_0^2}\frac 1{\Q^2}\\
&&\ds\quad\times k_0^2\paq{k_0^2\delta_{ik}-\pa{k+q}_ik_k -q_iq_k}\paq{k_0^2\delta_{jl}-\pa{k+q}_jk_l-q_jq_l}\,,
\ea
\ee
which exactly matches the sum of gravitational (\ref{eq:EQQ_GR}) and dilatonic
(\ref{eq:EQQ_psi}) electric quadrupole tail self-energies at $O(G_N^2)$. 

The hereditary (non-local in time) structure of this diagram comes from the terms
$\sim k^6_0 I_{ij}^2$, which displays a divergence from the integration region
$\Q\to 0$, $\K\to\infty$.
After the $\K,\Q$ integration a logarithmic piece of the type
$\int {\rm d}k_0k_0^6\log(k_0)I_{ij}^2$ is obtained, that is local in $k_0$-space but
translates in direct space into
$\int_{-\infty}^\infty {\rm d}t\, I_{ij}(t)\int_0^\infty \frac{{\rm d}\tau}\tau I_{ij}^{(6)}(t-\tau)$
which is the long-known hereditary term \cite{BD88,Foffa:2011np}.

Generalization to higher order electric multipole moment is straightforward and
gives for the gravity+dilaton process:

\be
\label{eq:EOO_GRpsi}
\ba{rcl}
\ds {\cal S}_{GR+\psi}^{(EI^2_{r+2})}&=&\ds
\frac{E}{16\paq{\pa{r+2}!}^2\Lambda^4}\int\frac{{\rm d}k_0}{2\pi}
I^{iji_1\cdots i_r}(k_0)I^{klk_1\cdots k_r}(-k_0)\\
&&\ds\int_{\K,\Q}\frac{k_{i_1}\cdots k_{i_r}}{\K^2-k_0^2}
     \frac{\pa{k+q}_{k_1}\cdots\pa{k+q}_{k_r}}{(\K+\Q)^2-k_0^2}\frac 1{\Q^2}\\
     &&\ds\quad\times
     k_0^2\Big\{\pa{k_0^2\delta_{ik}-\pa{k+q}_ik_k-q_iq_k}\pa{k_0^2\delta_{jl}-\pa{k+q}_jk_l-q_jq_l}\\
&&\ds \quad +\delta_{jl}\paq{2k_0^2\pa{q_ik_k+q_k \pa{k+q}_i}-
k_i\pa{k+q}_k\pa{\pa{\K^2-k_0^2}+\pa{\pa{\K+\Q}^2-k_0^2}-\Q^2}}\\
&&\quad+\paq{k_ik_jq_kq_l-q_iq_j\pa{k+q}_k\pa{k+q}_l}\Big\}\,,
     \ea
     \ee
which matches the double copy of the gauge amplitude     
\be
\label{eq:qddh_A}
\ba{rcl}
\ds {\mathcal S}_{A}^{(qd^2_{r+1})}&=&\ds
-\frac{g^4q^a}{\paq{\pa{r+1}!}^2}\int\frac{{\rm d}k_0}{2\pi}d^{b,ii_1\cdots i_r}(k_0)d^{c,kk_1\cdots k_r}(-k_0)\pa{if^{abc}}\\
&&\ds\quad\times\int_{\K,\Q}\frac 1{\K^2-k_0^2}\frac 1{(\K+\Q)^2-k_0^2}\frac 1{\Q^2}\\
&&\ds\qquad \times k_{i_1}\cdots k_{i_r}\pa{k+q}_{k_1}\cdots\pa{k+q}_{k_{r}}
k_0\paq{k_0^2\delta_{ik}-\pa{k+q}_ik_k-q_iq_k}\,,
\ea
\ee
using the correspondence dictionary already established and the fact that the
last two lines of eq.~(\ref{eq:EOO_GRpsi}) vanish, as demonstrated in app.~\ref{sec:dc_show}.

\subsection{Tail diagrams with magnetic moments}

Analogously, for the tail of the magnetic quadrupole one can compute the contribution to the self-energy at $O(G_N^2)$ from the purely gravitational sector (no-dilaton involved at any vertex)
\cite{Foffa:2019eeb}
\be
\label{eq:EJJ_GR}
\ba{rcl}
\ds {\cal S}_{GR}^{(EJ^2)}&=&\ds
\frac E{9\Lambda^4}\int \frac{{\rm d}k_0}{2\pi}J^{ij}(k_0)J^{kl}(-k_0)
\int_{\K,\Q}\frac 1{\K^2-k_0^2}\frac 1{\pa{\K+\Q}^2-k_0^2}\frac 1{\Q^2}\times\\
&&\ds\left\{
\frac{k_0^4}8\paq{\delta_{jl}\pa{\delta_{ik}\K\cdot\pa{\K+\Q}-k_k\pa{k+q}_i}
  +\epsilon_{ilr}\epsilon_{kjs}k^r\pa{k+q}^s}\right.\\
&&\ds+\frac{k_0^2}4k_j\paq{q_l\pa{\delta_{ik}\K\cdot\pa{\K+\Q}-k_k\pa{k+q}_i}-
  \epsilon_{iln}\epsilon_{krs}k^nk^rq^s}\\
&&\ds\left.
     -\frac 18k_j\pa{k+q}_l\Big[
       \K\cdot(\K+\Q)\pa{\delta_{ik}\K\cdot\pa{\K+\Q}-k_k\pa{k+q}_i}
       +\epsilon_{imn}\epsilon_{krs}k^mq^nk^rq^s\Big]\right\}\,,
\ea
\ee
to which the axionic contribution only must be added.
The axion couples to both the dilaton and the gravity field $\phi$ (coupling in the last two lines of eq.~(\ref{eq:sbulk_cubic})),
but the computation can be simplified by observing that the process involving a
$\psi$ exactly cancels the process involving the Lagrangian where $\phi$
couples to $H_{\mu\nu\rho}^2$, with the only contribution coming from
the coupling $\phi H_{ij0}^2$, see eq.~(\ref{eq:sbulk_cubic}). In summary one
gets
\be
\label{eq:EJJ_B}
\ba{rcl}
\ds {\cal S}_{B}^{(EJ^2)}&=&\ds
\frac E{72\Lambda^4}\int \frac{{\rm d}k_0}{2\pi}J^{ij}(k_0)J^{kl}(-k_0)
\int_{\K,\Q}\frac 1{\K^2-k_0^2}\frac 1{\pa{\K+\Q}^2-k_0^2}\frac 1{\Q^2}\\
&&\ds\times\left\{
k_0^4\paq{\delta_{jl}\pa{\delta_{ik}\K\cdot\pa{\K+\Q}-k_k\pa{k+q}_i}-
  \epsilon_{ilr}\epsilon_{kjs}k^r\pa{k+q}^s}\right.\\
&&\ds +k_0^2\paq{-\pa{\pa{k+q}_j\pa{k+q}_l+k_jk_l}
  \pa{\delta_{ik}\K\cdot\pa{\K+\Q}-k_k\pa{k+q}_i}
  +2k_j\epsilon_{iln}\epsilon_{krs}k^nk^rq^s}\\
&&\ds\left.+k_j(k+q)_l\Big[\K\cdot(\K+\Q)
  \pa{\delta_{ik}\K\cdot(\K+\Q)-k_k\pa{k+q}_i}
  +\epsilon_{imn}\epsilon_{krs}k^mq^nk^rq^s\Big]\right\}\,.
\ea
\ee
For the magnetic quadrupole, like in the leading order self-energy, all
terms with the Levi-Civita tensors cancel when adding the gravity and axionic
contributions, to give
\be
\label{eq:EJJ_GRB}
\ba{rcl}
\ds S_{GR+B}^{(EJ^2)}&=&\ds
\frac E{36\Lambda^4}\int \frac{{\rm d}k_0}{2\pi}J^{ij}(k_0)J^{kl}(-k_0)
\int_{\K,\Q}\frac 1{\K^2-k_0^2}\frac 1{\pa{\K+\Q}^2-k_0^2}\frac 1{\Q^2}\\
&&\ds \times \left\{k_0^2\Big[\delta_{ik}\K\cdot\pa{\K+\Q}-\pa{k+q}_ik_k\Big]
\paq{k_0^2\delta_{jl}-k_l\pa{k+q}_j-q_jq_l}\right.\\
&&\ds\left.\quad +\frac{k_0^2}2\paq{\delta_{ik}\K\cdot\pa{\K+\Q}-\pa{k+q}_ik_k}
\paq{k_j q_l+q_j (k+q)_l}\right\}\,,
\ea
\ee
which indicates a factorizable structure, even though not a perfect
square, by observing that the integral of the last line vanishes, see explicit
computations in app.~\ref{sec:dc_show}.

Computing the magnetic dipole tail diagram on the gauge side one gets
\be
\label{eq:qmm_gauge}
\ba{rcl}
\ds {\mathcal S}_{A}^{(q\mu^2)}&=&\ds i\frac{q^ag^4}8\int \frac{{\rm d}k_0}{2\pi}\mu^{bi}(k_0)\mu^{ck}(-k_0)\pa{if^{def}}
\epsilon_{imn}\epsilon_{krs}\\
&&\ds\quad\times\int_{\K,\Q}\frac 1{\K^2-k_0^2}\frac 1{\pa{\K^2+\Q^2}-k_0^2}\frac 1{\Q^2}
%&&\qquad \times
\langle\pa{A_0^aF^b_{\K mn}F_{-\K-\Q rs}^c}\pa{\partial_\mu A_\nu^d A_\rho^e A_\sigma^f}\rangle'
\eta^{\mu\rho}\eta^{\nu\sigma}\\
&=&\ds -q^ag^4\int \frac{{\rm d}k_0}{2\pi}\mu^{bi}(k_0)\mu^{ck}(-k_0)\pa{if^{abc}}\int_{\K,\Q}
\frac 1{\K^2-k_0^2}\frac 1{\pa{\K+\Q}^2-k_0^2}\frac 1{\Q^2}\\
&&\ds \qquad\times k_0\pa{\delta_{ik}\K\cdot\pa{\K+\Q}-k_k\pa{k+q}_i}\,.
\ea
\ee
Much like in the leading order self-energy for magnetic sources,
to reproduce the gravitational plus axionic magnetic quadrupole tail one
should not square (\ref{eq:qmm_gauge}), but rather combine it with the electric tail
eq.~(\ref{eq:qdd_A}), according to previously derived rules (\ref{eq:rule_d})
, (\ref{eq:rule_m}) and (\ref{eq:rule_qdd}) complemented with
\be
\frac 12\epsilon_{imn}\langle F_{\K mn}^aA_\mu^b\rangle'=
i\epsilon_{im\mu}k_m\delta^{ab}\to -\epsilon_{im\mu}k_m\pa{k_j\eta_{0\nu}-k_0\eta_{j\nu}}\,,
\ee
which is consistent with replacing $\delta^{ab}$ with a contraction between a
gauge field and an electric field $\sim \langle F_{\K j0}A_\nu\rangle'$.
One then obtains
\be
\ba{rcl}
   {\cal S}_{DC}^{(EJ^2)}&=&\ds\frac E{36\Lambda^4}\int \frac{{\rm d}k_0}{2\pi}J^{ij}(k_0)J^{kl}(-k_0)
   \int_{\K,\Q}\frac 1{\K-k_0^2}\frac 1{\pa{\K+\Q}^2-k_0^2}\frac 1{\Q^2}\times\\
&&\ds \paq{k_0^2\pa{\delta_{ik}\K\cdot\pa{\K+\Q}-k_k\pa{k+q}_i}\pa{k_0^2\delta_{jl}-k_l\pa{k+q}_j-q_jq_l}}\,,
   \ea
   \ee
   which matches (\ref{eq:EJJ_GRB}).
   
Finally we give the formula for the double copy in the case of higher order magnetic moments
$J^{iji_1\cdots i_r}$ for gravity+axion exchange:
\be
\label{eq:EMM_DC}
\ba{rcl}
\ds{\cal S}^{(EJ^2_{r+2})}_{DC}&=&\ds\frac E{9\paq{\pa{r+2}!}^2\Lambda^4}
\int\frac{{\rm d}k_0}{2\pi}J^{iji_1\dots i_r}(k_0)\,J^{klk_1\dots k_r}(-k_0)\\
&&\ds\intkq\, \frac 1{\K^2-k_0^2}\frac 1{\pa{\K+\Q}^2-k_0^2}\frac 1{\Q^2}
k_{i_1}\cdots k_{i_r}(k+q)_{k_1}\cdots (k+q)_{k_r}\\
&&\ds\times\paq{k_0^2\pa{\delta_{ik}\K\cdot\pa{\K+\Q}-k_k\pa{k+q}_i}
  \pa{k_0^2\delta_{jl}-k_l\pa{k+q}_j-q_jq_l}}\,.
\ea
\ee

\section{Discussion}
\label{sec:discussion}
The question of how the classical two-body gravitational potential may be extracted
from quantum scattering amplitudes has a long history and investigation has
been revived recently by works extending the class of double-copy applications to
perturbative solutions of the equations of motions and to the effective
action of a binary system. Our investigations aim at providing further
evidence that the classical double copy can be applied to the derivation of
gravitational two-body potential, which is relevant for theoretical modeling
of gravitational wave sources.

In particular the post-Newtonian approach has been useful in constructing
templates for gravitational wave data analysis and it decomposes the
problem into a \emph{near} and a \emph{far} zone, the former involving longitudinal modes
only, whereas the latter includes both longitudinal and radiative gravitational modes.
Focusing on the far zone, where the gravitational wave source is defined
as an extended object with multipoles, we have shown how the next-to-leading
order in the Newton's constant responsible for tail terms in the effective
potential
can be reproduced with a double-copy procedure, applied for the first time to
the \emph{multipolar} post-Minkowskian expansion.

Future applications to post-Newtonian, post-Minkowskian and multipolar
approximations include application to $O(G_N^2)$ self-energy processes contributing to
the effective action under the name of memory effect, and a systematization
to higher order will be necessary.
Note however that care is needed when comparing $G_N$ order among different approaches:
in our case, for instance, we have terms $G_N^2\dddot I_{ij}^2$ that when expressed in terms
of individual binary system constituent kinematic variables, involve acceleration and higher derivatives
which can be expressed in terms of position and velocity via equations of motion.
The tail term, studied in sec.~\ref{sec:self_energy_nlo}, is e.g. a fourth post-Newtonian term giving rise
to $G_N^4$ terms in the equations of motion.

\section*{Acknowledgments}
The work of R.S. is partly supported by CNPq.
The research of R.S. was partly supported by ICTP-SAIFR, by the International Centre for
Theoretical Sciences (ICTS) during a visit for participating in the program -
The Future of Gravitational-Wave Astronomy (Code: ICTS/fgwa2019/08)
and by the Munich Institute for Astro- and Particle Physics (MIAPP) which is
funded by the Deutsche Forschungsgemeinschaft (DFG, German Research Foundation)
under Germany's Excellence Strategy – EXC-2094 – 390783311. G.L.A. thanks the research funding agency CAPES for the Ph.D. scholarship.
S.F. is supported by the Fonds National Suisse and by the SwissMap NCCR.

\appendix

\section{Computation details}
\label{app:details}
\subsection{Multipole expansion in dilaton-axion-gravity}

The mutipole expansion is obtained by Taylor expanding the terms
bilinear in the source and the gravitational-dilatonic-axionic field,
and collecting terms at the same order of $v$, being $v$ the typical internal
velocity of the source.

Describing the source as a continuous extended body (instead of the equivalent
description of a collection of point particle used in sec.~\ref{sec:method}),
one can characterize the source with its energy momentum tensor $T^{\mu\nu}$ and
spin density $s^{\mu\nu}$ extended over a volume $V$:
\be
   {\cal S}_{source}=\int {\rm d}t\int_V{\rm d}^3x\pa{\frac 12T^{\mu\nu}h_{\mu\nu}+T\dfrac \psi\Lambda+
\frac1{4\Lambda} \dot x^\rho s^{\mu\nu}H_{\mu\nu\rho}}\,.
\ee
Considering that the source is localized in a region or size $r$ much smaller
than the radiation wavelength $\lambda_r\sim 2\pi/\omega\sim r/v$ one can Taylor expand
${\cal S}_{source}$ to obtain
\be
\ba{rcl}
\ds {\cal S}_{mult}&\simeq&\ds\int {\rm d}t\left\{\pa{\int_V {\rm d}^3xT^{00}}h_{00}+
\paq{2 \pa{\int_V {\rm d}^3xT^{0i}}h_{0i}+\pa{\int_V {\rm d}^3x T^{00}x^i}h_{00,i}}\right.\\
&&\ds+\paq{
  \pa{\int_V {\rm d}^3xT^{ij}}h_{ij}+\pa{\int_V {\rm d}^3x T^{0i}x^j}\pa{h_{0i,j}+h_{0j,i}}+
  \frac 12\pa{\int_V {\rm d}^3x T^{00}x^ix^j}h_{00,ij}}\\
&&\ds\left.\pa{\int_V {\rm d}^3x T^{0i}x^j}\pa{h_{0i,j}-h_{0j,i}}+
\pa{\int_V {\rm d}^3x T^{ij}x^k}h_{ij,k}+\ldots\right\}\,.
  \ea
  \ee
  Note that this Taylor expansion is actually an expansion in $r/\lambda_r\sim v$. 
  Using repeatedly the energy-momentum conservation in the form $\dot T^{\mu 0}=-T^{\mu i}_{\ \ ,i}$
  one can derive
  \be
  \ba{rcl}
\ds  \int_V {\rm d}^3x T^{ij}&=&\ds\frac 12\frac {\rm d}{{\rm d}t}\pa{\int_V {\rm d}^3x T^{00}x^ix^j}\equiv \frac 12\ddot{I}^{ij}\,,\\
\ds  \int_V {\rm d}^3x T^{0i}&=&\ds -\frac {\rm d}{{\rm d}t}\pa{\int_V {\rm d}^3xT^{00}x^i}\,,\\
\ds  \int {\rm d}^3x T^{ij}x^k&=&\ds\frac 13\int {\rm d}^3x\pa{T^{ij}x^k+T^{ki}x^j+T^{jk}x^i}+\frac 13\int {\rm d}^3x\pa{2T^{ij}x^k-T^{ik}x^j-T^{jk}x^i}\,,\\
&=&\ds\frac 16\frac{{\rm d}^2}{{\rm d}t^2}\pa{\int {\rm d}^3x T^{00}x^ix^jx^k}
+\frac 13\frac {\rm d}{{\rm d}t}\paq{\int {\rm d}^3x \pa{T^{0i}x^kx^j+T^{0j}x^kx^i-2T^{0k}x^ix^j}}\,.
\ea
\ee
Hence the electric quadrupole coupling $T^{ij}h_{ij}$ give rise to the
$\frac 12\ddot Q^{ij}R_{0i0j}$ term,
where at linear order in terms of the Kaluza-Klein fields $\phi,A_i,\sigma_{ij}$
\be
R_{0i0j}\simeq \frac 12\pa{\ddot \sigma_{ij}-\dot A_{i,j}-\dot A_{j,i}-2\phi_{,ij}-2\delta_{ij}\frac{\ddot\phi}{d-2}}+O(h^2)\,,
\ee
and using
\be
\ba{l}
\ds\pa{T^{0i}x^kx^j-T^{0k}x^ix^j}\sigma_{ij,k}=
T^{0m}x^nx^j\pa{\delta^l_m\delta^k_n-\delta^k_m\delta^l_n}\sigma_{lj,k}\\
\ds =-\epsilon_{imn}x^mT^{0n}x^j\frac 12\epsilon^{ikl}\pa{\sigma_{kj,l}-\sigma_{lj,k}}
\ea
\ee
one finds the gravitational magnetic quadrupole coupling eq.~(\ref{eq:mult})
using the definition of the magnetic part of the Riemann tensor
\be
   {\cal B}_{ij}=\frac 12\epsilon_{ikl}R_{0jkl}\simeq \frac 1{4\Lambda}
   \epsilon_{ikl}\paq{\dot \sigma_{jk,l}-\dot\sigma_{jl,k}+ A_{l,jk}-A_{k,jl}+
     \frac 2{d-2}\pa{\dot\phi_{,k}\delta_{jl}-\dot\phi_{,l}\delta_{jk}}}\,.
\ee

Analogously for the axion field one has
\be
\ba{l}
\ds \int {\rm d}t\paq{\pa{\int_V {\rm d}^3x s^{ij}}H_{0ij}+\pa{\int_V {\rm d}^3x s^{ij}x^k}H_{0ij,k}+
\pa{\int_V{\rm d}^3x s^{ij}v^k}H_{ijk}+O(v^2)}\\
\ds =\int {\rm d}t\paq{ \pa{\int_V {\rm d}^3x s^{ij}}H_{0ij}+\pa{\int_V {\rm d}^3x s^{ij}x^k}
\pa{H_{ij0,k}-\dot H_{ijk}}+O(v^2)}\,,
\ea
\ee
where integration by part has been used and terms involving
$s^{0i}\simeq s^{ij}v_j$ have been neglected since they enter at order $v^2$ with
respect to the leading one.
Finally introducing the spin density pseudo-vector $\tilde s^i$ dual to
$s^{ij}$ one has for the coupling of the first moment of the spin coupling to the axion
\be
\ba{l}
\ds\frac 12\epsilon^{ijl}\int_V\pa{\tilde s^lx^k+\tilde s^kx^l+\tilde s^lx^k-\tilde s^kx^l}\pa{H_{ij0,k}-\dot H_{ijk}}\\
\ds=\frac 43J^{kl}\epsilon^{ijl}\pa{B_{0i,k}+\dot B_{ik}+B_{k0,i}}_{,j}+
\frac 12\epsilon^{ijl}\int_V\pa{\tilde s^lx^k-\tilde s^kx^l}\pa{B_{0i,k}+\dot B_{ik}}\,,
\ea
\ee
where it has been used that the leading spin contribution to the magnetic
quadrupole is (the traceless part of) $\frac 34\pa{\tilde s^lx^k+\tilde s^kx^l}$
\cite{Blanchet:2006gy}, thus recovering the magnetic quadrupole coupling
to the axion in eq. (\ref{eq:mult}), beside a coupling to the antisymmetric
first moment of the spin which has no gravitational analog.

\subsection{Graviton, dilaton, axion action up to cubic interaction}
The bulk action is needed for the computations of this paper up to cubic
interaction in gravitational, dilatonic and axionic field and it is reported here
explictly:
\renewcommand{\arraystretch}{1.4}
\be
\label{eq:sbulk_cubic}
\ba{rcl}
\ds {\mathcal S}_{bulk} &\supset &\ds \int {\rm d}^{d+1}x\sqrt{-\gamma}
\left\{\frac{1}{4}\left[(\vec{\nabla}\sigma)^2-2(\vec{\nabla}\sigma_{ij})^2-\left(\dot{\sigma}^2-2(\dot{\sigma}_{ij})^2\right){\rm e}^{\frac{-c_d \phi}{\Lambda}}\right]- c_d \left[(\vec{\nabla}\phi)^2-\dot{\phi}^2 {\rm e}^{-\frac{c_d\phi}{\Lambda}}\right]\right.\\
&&\ds\quad
+\paq{\frac{F_{ij}^2}{2}+\left(\vec{\nabla}\!\!\cdot\!\!\vec{A}\right)^2 -\dot{\vec{A}}^2 {\rm e}^{-\frac{c_d\phi}{\Lambda}}}{\rm e}^{\frac{c_d \phi}{\Lambda}}+
\frac 2\Lambda\paq{\pa{F_{ij}A^i\dot{A^j}+\vec{A}\!\!\cdot\!\!\dot{\vec{A}}(\vec{\nabla}\!\!\cdot\!\!\vec{A})}-c_d\dot{\phi}\vec{A}\!\!\cdot\!\!\vec{\nabla}\phi}\\
&&\ds\quad
-\frac{1}{\Lambda}\pa{\frac{\sigma}{2}\delta^{ij}-\sigma^{ij}}
\pa{{\sigma_{ik}}^{,l}{\sigma_{jl}}^{,k}-{\sigma_{ik}}^{,k}{\sigma_{jl}}^{,l}+\sigma_{,i}{\sigma_{jk}}^{,k}-\sigma_{ik,j}\sigma^{,k}}\\
&&\ds\quad
+\frac{\dot{\sigma}_{ij}}{\Lambda}\pa{-\delta^{ij}A_l\hat{\Gamma}^l_{kk}+ 2A_k\hat{\Gamma}^k_{ij}-2A^i\hat{\Gamma}^j_{kk}}\\
&&\ds\quad +2(d-1)\paq{e^{-c_d\frac\phi\Lambda}\dot\psi^2-(\nabla\psi)^2-
  \frac 2\Lambda \dot\psi\vec A\cdot\vec\nabla\psi}\\
&&\ds\quad\left. -\frac 16e^{\frac{4\psi}\Lambda+2(c_d-2)\frac\phi\Lambda}\pa{H_{ijk}^2-3e^{-c_d\frac\phi\Lambda}H_{ij0}^2}-A^iH_{ijk}H_{0jk}+\frac 12H_{ijk}H_{ijl}\sigma^{kl}\right\}\,.
\ea
\ee

Contractions between explicit space (Latin) indices are done with flat metric,
when indices are understood contractions are made including the field
$\sigma_{ij}$ in the metric, e.g.
$\vec A\cdot \nabla\phi=A_i\phi_{,j}\pa{\delta^{ij}-\sigma^{ij}+\ldots}$.

\section{Vanishing integrals}
\label{sec:dc_show}

To recast the gravitational amplitudes in eqs.~(\ref{eq:MEE_DC}) and (\ref{eq:EOO_GRpsi})
into a double-copy structure, we now show that the pieces that do not fit the factorizable
form vanish identically.
Performing the change of integration variables $\Q\to -\pa{\K+\K'}$
one immediately finds:
\be
\ba{l}
\ds\int_{\K,\Q}\frac{k_{i_1}\cdots k_{i_n}\pa{k+q}_{k_1}\cdots\pa{k+q}_{k_n}}
       {\pa{\K^2-k_0^2}\pa{\pa{\K+\Q}^2-k_0^2}\Q^2}
       \pa{q_ik_k+q_k\pa{k+q}_i}\\
       \ds
= \pa{-1}^n\int_{\K,\K'}\frac{k_{i_1}\cdots k_{i_n}k'_{k_1}\cdots k'_{k_n}}
           {\pa{\K^2-k_0^2}\pa{{\K'}^2-k_0^2}\pa{\K+\K'}^2}
           \paq{k'_ik'_k-k_ik_k}\,,
\ea
\ee
which vanishes when contracted with $\delta_{jl}I^{iji_1\cdots i_n}(k_0)I^{klk_1\cdots k_n}(-k_0)$,
being antisymmetric under swapping $k\leftrightarrow k'$.

Terms proportional to gravitational radiation propagators $(\K^2-k_0)$
or $[\pa{\K+\Q}^2-k_0^2]$ also vanish identically, given that 
\be
\int_\Q\frac{q_{i_1}\cdots q_{i_n}}{\Q^2}=0
\ee
for any $n$.

For terms proportional to $\Q^2$ we observe that
\be
\int_{\K,\K'}\frac{k_{i_1}\cdots k_{i_{2n}}k'_{k_1}\cdots k'_{k_{2m}}}{\pa{\K^2-k_0^2}\pa{{\K'}^2-k_0^2}}\propto \delta^{(i_1i_2}\cdots\delta^{i_{2n-1}i_{2n})}\delta^{(k_1k_2}\cdots\delta^{k_{2m-1}k_{2m})}\,
\ee
vanish when contracted with traceless tensors, and also trivially
vanish when an odd number of momentum factors in the numerator is involved.

Finally by rearranging the last term as follows
\be
\ba{l}
\ds\intkq \frac{k_{i_1}\cdots k_{i_n}\pa{k+q}_{k_1}\cdots\pa{k+q}_{k_n}}
          {\pa{\K^2-k_0^2}\pa{\pa{\K+\Q}^2-k_0^2}\Q^2}\times\pa{
            k_ik_jq_kq_l-q_iq_j\pa{k+q}_k\pa{k+q}_l}\\
\ds=(-1)^n\int_{\K,\K'}\frac{k_{i_1}\cdots k_{i_n}k'_{k_1}\cdots k'_{k_n}}
    {\pa{\K^2-k_0^2}\pa{{\K'}^2-k_0^2}\pa{\K+\K'}^2}\\
\ds\qquad\times
    \paq{k_ik_jk_kk_l-k'_ik'_jk'_kk'_l+k_ik'_l\pa{k_jk_k-k'_jk'_k}+
      k_jk'_k\pa{k_ik_l-k'_ik'_l}}\,,
\ea
\ee
one sees that it vanishes when contracted with $I^{iji_1\cdots i_n}I^{klk_1\cdots k_n}$
because of the anti-symmetry under $k\leftrightarrow k'$.
This concludes the demonstration that terms in eqs.~(\ref{eq:MEE_DC}) and
(\ref{eq:EOO_GRpsi}) that do not fit in the double copy
structure vanish. Last line in eq.~(\ref{eq:EJJ_GRB}) can be shown to vanish
with the same reasoning, using momenta $\K$ and $\K'\equiv -(\K+\Q)$.

\end{document}